\documentclass{llncs}
\usepackage{soul}

\title{Concept Embedding for Information Retrieval}
\titlerunning{Concept Embedding for Information Retrieval}
\author{Karam Abdulahhad}
\institute{GESIS - Leibniz Institute for the Social Sciences, Cologne, Germany
\email{karam.abdulahhad@gesis.org}}
\authorrunning{K. Abdulahhad}

\begin{document}

\setlength{\abovedisplayskip}{2pt}
\setlength{\belowdisplayskip}{2pt}

\maketitle

\begin{abstract}
    Concepts are used to solve the term-mismatch problem. However, we need an effective similarity measure between concepts. Word embedding presents a promising solution. 
    We present in this study three approaches to build concepts vectors based on words vectors. We use a vector-based measure to estimate inter-concepts similarity. Our experiments show promising results. Furthermore, words and concepts become comparable. This could be used to improve conceptual indexing process.
\end{abstract}

\section{Introduction}

Conceptual indexing includes the process of annotating raw text by concepts\footnote{Concepts have many definitions \cite{ka2014}. A concept here refers to a category ID that encompasses synonymous words and phrases, e.g. UMLS concepts, WordNet synsets.} of a particular knowledge source \cite{ka2014}. It is used to represent the content of documents and queries by more \emph{informative} terms, namely concepts rather than words. Annotating text by concepts is used to solve the term-mismatch problem by considering the semantic of text rather than its form \cite{ka2014}. For example, the two terms ``cancer'' and ``malignant neoplastic disease'' correspond to the same concept (\emph{synset}) in WordNet\footnote{wordnet.princeton.edu}. However, using concepts instead of words has some side-effects. First, the process of annotating text by concepts is a potential source of noise, e.g. ``x-ray'' corresponds to more than 6 different concepts in UMLS\footnote{www.nlm.nih.gov/research/umls/}, so which one best fits the original textual content. Second, 
to better solve the term-mismatch problem, we need to exploit the relations between concepts. Hence, we need a way to quantify these relations. 
However, inter-concepts similarity is still problematic and non easy to measure \cite{tpx2007}, because the similarity between two concepts depends on the relation between them. Since relations are different in semantic, e.g. \emph{is-a}, \emph{part-of}, etc., and have different properties, e.g. symmetric or not, there is no standard way on how to quantify relations, where it is task-dependent. For example, for one task the \emph{is-a} relation is much useful than \emph{part-of}, but for another task the \emph{part-of} is much useful, and so on.

Word embedding \cite{tmx2013nips,jpx2014} has recently proved its effectiveness for several NLP tasks. It is also studied in Information Retrieval (IR), where word embedding is used for ad-hoc retrieval \cite{jgx2016a}, query expansion \cite{gzx2015a,fdx2016}, or text similarity \cite{tkx2015}. Some features make word embedding potentially useful for IR, where a word is a low-dimensional numerical vector rather than a sequence of characters and algebraic operations between vectors reflect semantic relatedness between words \cite{tmx2013nips}.

Concept embedding takes word embedding to a higher level. It is the process of representing concepts by low-dimensional vectors of real numbers. Through concept embedding, one can keep the advantages of both conceptual indexing and word embedding, and at the same time avoid some of conceptual indexing disadvantages. More precisely, by using concepts vectors, on the one hand, we exploit concepts to reduce the term-mismatch effect, and on the other hand, we avoid complexities related to relation-based inter-concept similarity, and measure the similarity between two concepts by comparing their corresponding vectors.

In this study, we propose a way to generate concept embedding based on word embedding. Then, we use concepts vectors in classical IR models. It is worth mentioning that we do not aim in this study to compare different approaches of tackling term-mismatch. Hence, we do not report in results any comparison between our approach and approaches like: pseudo-relevance feedback or word based expansion. The main goal of this paper is to check the \emph{profitability of using concept embedding, and the adaptability of vector based concept similarity to IR}.


\section{Related Works}

De Vine et al. \cite{ldvx2014} build medical concept embedding through replacing the textual content of documents by their corresponding medical concepts, and then training \emph{word2vec} \cite{tmx2013nips} on the new corpus, which is now a sequence of concepts. At the end of the process, they obtain a vector representation for each concept appeared in the corpus. Choi et al. \cite{ycx2016} use a similar approach to obtain concepts vectors, except that they use temporal information from medical claims to adapt the definition of context window of \emph{word2vec} to medical data. Both approaches build vectors for the concepts that only appear in the corpus and not for all concepts of the corresponding knowledge resources. Furthermore, if we build word embedding vectors of the same corpus, then concepts vectors and words vectors will not be comparable, because they are represented in different vector spaces.

Several studies proposed to use word embedding to represent more informative elements rather than a single word. Clinchant et al. \cite{scx2013a} use Fisher kernel to aggregate words vectors of a document to build a document vector. Le et al. \cite{qlx2014} extend \emph{word2vec} to be able to compute paragraph-level embedding. 
Zamani et al. \cite{hzx2016a} optimize a query language model to estimate the embedding vector of a query, where averaging query's words vectors is a special case of their approach.

Concerning the inter-concepts similarity, many approaches have been used in literature \cite{tpx2007}. They can be categorized \cite{tpx2007}: 1- \emph{path-based} measures, which depend on the length and the nature of the path that links two concepts within a knowledge resource; 2- \emph{information content} measures, which use some corpus-based statistics to estimate the information content of a concept, and then measuring similarity; and 3- \emph{vectors-based} measures, which depend on the ability to represent concepts by vectors, where $\cos$ is the main measure in this category.


\section{Concept Embedding}

We present in this study three methods for concept embedding based on word embedding. The difference between these methods is the additional information that is used, beside word embedding vectors, to build concepts vectors.




\emph{\textbf{Flat embedding (FEmb)}}: In this method, we do not use any additional information rather than word embedding vectors. The main hypothesis here is that any concept can be mapped to a set of words. Hence, the embedding vector of a concept $c$ is a function $F$ of the vectors of its words. For example, in WordNet the two words ``snake'' and ``serpent'' belong to the ``S01729333'' synset, so $\overrightarrow{\mathit{S01729333}} = F(\overrightarrow{\mathit{snake}}, \overrightarrow{\mathit{serpent}})$, where $F$ is any function able to merge several vectors in only one vector, e.g. vectors addition, vectors average, etc.


\emph{\textbf{Hierarchical embedding (HEmb)}}: Beside word embedding vectors, we use in this method the internal structural information of each concept. This method is initially proposed to deal with UMLS medical concepts, but it is applicable to any resource exhibiting similar concept structure. In UMLS, each \emph{concept} $c$ consists of several \emph{terms}, which represent the different forms of text that could be used to express the underlying meaning of $c$, and each term could appear in different lexical variations or \emph{strings}, where a string can be mapped to either a word or a set of words. Therefore, we have a hierarchy related to each concept. Assume that a concept $c$ consisting of two terms $t_1, t_2$, and each term $t_i$ consists of two strings $s_1^i, s_2^i$. In this case, $\overrightarrow{c} = F\left(F\left(s_1^1, s_2^1\right),F\left(s_1^2, s_2^2\right)\right)$.

\emph{\textbf{Weighted embedding (WEmb)}}: This method is an extension of \emph{FEmb}, where we incorporate external statistical information. More precisely, instead of equally treating the words of each concept, we attach a weight indicating their relative importance, i.e. $\overrightarrow{c} = F(\alpha_1 \overrightarrow{w_1},\dots,\alpha_n \overrightarrow{w_n})$, where $\alpha_i$ is the weight of $w_i$.

\emph{\textbf{Evaluation strategy}}: Since our goal is to study the profitability of concept embedding for IR, we evaluate the retrieval performance improvement of an IR model that is able to incorporate inter-terms similarity. We use the model of \cite{fc2000}:
%
\begin{equation}
\label{ir}
    RSV(d,q) = \sum_{c \in q} weight_q(c) \times sim(c,c^*) \times weight_d(c^*)
\end{equation}
where $sim(c,c^*)$ is the similarity between two concepts, and $c^*$ is the closest document concept to the query concept $c$ according to the similarity measure $sim$. If the query concept $c$ also belongs to $d$, then $c^* = c$ and $sim(c,c^*)=1$. We use several definitions for $sim$, some of them are vector based and some are not. By this way we can see if concept embedding vectors are useful for IR.

\section{Experimental Setup}


\emph{\textbf{Generating word embedding}}: We generate concept embedding vectors based on word embedding vectors. To obtain words vectors, we train \emph{word2vec} on open access \emph{PubMed Central} collection\footnote{www.ncbi.nlm.nih.gov/pmc/, \emph{PubMed} collection contains: 1177879 vocabularies.}, with the following configurations: vector size 500, continuous bag of words, window size 8, and negative sampling is set to 25.

\emph{\textbf{Generating concept embedding}}: We apply our approach to \emph{UMLS2017AA} medical concepts, and we only consider the concepts that have English content. 
Assume the following example for clarification. The concept \emph{C0004238} (denoted $c$) has two textual forms or terms: \emph{L0004238} (denoted $t_1$) and \emph{L0004327} (denoted $t_2$). Term $t_1$ appears in two lexical variations: singular \emph{S0016668=``atrial fibrilliation''} (denoted $s^1_1$), and plural \emph{S0016669=``atrial fibrilliations''} (denoted $s^1_2$). The same for term $t_2$ which corresponds to two strings $s^2_1$ and $s^2_2$. By tokenizing, we transform each string $s^i_j$ to a set of words $W_{s^i_j}$ (\emph{we remove duplication}).

In \emph{\textbf{FEmb}}, the concept vector is: $\overrightarrow{c} = avg(\overrightarrow{w_1},\dots,\overrightarrow{w_l})$, where $\overrightarrow{w_i}$ is the word embedding vector of word $w_i$, $w_i \in \bigcup_{i,j}W_{s^i_j}$, and $avg$ returns the average of a set of vectors. For \emph{\textbf{HEmb}}, the concept vector is: $\overrightarrow{c} = avg(avg(t_1),\dots,avg(t_m))$, where $avg(t_i) = avg(avg(s^i_1),\dots,avg(s^i_k))$, $avg(s^i_j)=avg(\overrightarrow{w_1},\dots,\overrightarrow{w_l})$, and $w \in W_{s^i_j}$. Concerning \emph{\textbf{WEmb}}, we follow the same approach as \emph{FEmb}, except that we compute the weighted average $wavg$ instead of average $avg$. More precisely, $\overrightarrow{c} = wavg(\overrightarrow{w_1},\dots,\overrightarrow{w_l})=\frac{1}{l}\sum_{w}\alpha_w \overrightarrow{w}$, where $l$ is the number of words. The weight $\alpha_w$ of a word $w$ is its \emph{idf} score in \emph{PubMed}, namely $\alpha_w = \ln(\frac{N+1}{n})$, where $N$ is the number of documents in \emph{PubMed} and $n$ is document frequency of $w$.

We generate fixed random vectors for missing words, which means, if a missing word $w$ appears in several concepts, we use the same randomly generated vector. For the \emph{idf}-weight of missing words, we tested several options: assuming that the word is too popular ($n = N$), too rare ($n = 1$), or in between ($n = \frac{N}{2}$). The three approaches give similar performance; therefore, we only report the first option where $n=N$, which means, a poor \emph{idf} score.

\emph{\textbf{Test collections}}: To evaluate our proposal, we use ad-hoc image-based corpus of ImageCLEF (www.imageclef.org) of years 2011 (\emph{clef11}) and 2012 (\emph{clef12}), where documents are captions of medical images with short queries. \emph{clef11} has 230K documents and 30 queries. \emph{clef12} contains 300K documents and 21 queries (we removed query 14 because it is not mapped to any concept). Documents and queries are mapped to UMLS concepts using MetaMap (metamap.nlm.nih.gov).

\emph{\textbf{IR model and concept similarity}}: There are three components to be described in the IR model of (\ref{ir}). The weight of concepts in documents and queries, and the similarity between concepts. To compute the weight of a concept in a document or a query, we apply two classical IR weighting schema: Pivoted Normalization or BM25 \cite{ka2014}. For both models, we use standard parameters values reported in \cite{ka2014}. To compute the similarity between concepts $sim(c,c')$, we use two measures. The first one is compatible with the vector representation of concepts:
\begin{equation}
\label{sim}
    sim(c,c') =
        \left\{
            \begin{array}{ll}
                0 & \cos(\theta) \leq 0\\
                \beta \times \cos^2(\theta) \quad\quad & \mbox{otherwise}
            \end{array}
        \right.
\end{equation}
where $\theta$ is the angle between the two vectors $\overrightarrow{c}$ and $\overrightarrow{c'}$, and $\beta$ is a tuning parameter. 
We optimized the value of $\beta$ on \emph{clef11} but it is applied to all collections. In our results, we only report the retrieval performance of $\beta=0.5$. In addition, we only consider the similarity when $\cos(\theta) > 0$, i.e. we ignore the concepts that could have an opposite meaning. We use $\cos^2(\theta)$ instead of $\cos(\theta)$, because it is more discriminant, especially for small angles $\theta \in [-\frac{\pi}{4},\frac{\pi}{4}]$.
For comparison, we use Leacock measure \cite{tpx2007}, which depends on the length of the path of \emph{is-a} relations between two concepts in UMLS.

\section{Evaluation}

Table \ref{tab} shows results for \emph{clef11} and \emph{clef12}. 
\emph{FEmb}, \emph{HEmb}, and \emph{WEmb} refer to our approaches to build concepts vectors, where the similarity measure between concepts is (\ref{sim}). \emph{NoEmb} refer to deal with concepts rather than concepts vectors, where \emph{Leacock} refer to the similarity between concepts, whereas, we do not incorporate similarity in \emph{NoSim}. $*$ and $\dagger$ refer to a statistically significant difference with \emph{NoEmb\_NoSim} and \emph{NoEmb\_Leacock}, respectively, according to Fisher Randomization test with ($\alpha < 0.05$).
\vspace*{-5mm}
\begin{table}
    \caption{Experimental results for \emph{clef11} and \emph{clef12} collections}
    \vspace{-2mm}
    \centering
    \begin{tabular}{lllll|llll}
        \cline{2-9}
         & \multicolumn{4}{c|}{\emph{clef11}} & \multicolumn{4}{c}{\emph{clef12}}\\
        \cline{2-9}
         & \multicolumn{2}{c}{\emph{piv}} & \multicolumn{2}{c|}{\emph{bm25}} & \multicolumn{2}{c}{\emph{piv}} & \multicolumn{2}{c}{\emph{bm25}}\\
         & \emph{MAP} & \emph{P@10} & \emph{MAP} & \emph{P@10} & \emph{MAP} & \emph{P@10} & \emph{MAP} & \emph{P@10}\\
         \hline
         \hline
        \emph{NoEmb-NoSim} & 0.1096 & 0.2300 & 0.1552 & 0.3100 & 0.0978 & 0.1381 & 0.1083 & 0.1571\\
        \emph{NoEmb-Leacock} & 0.1085 & 0.2267 & 0.1505 & 0.2933 & 0.0927 & 0.1429 & 0.1064 & \textbf{0.1667}\\
        \hline
        \emph{FEmb\_Eq\ref{sim}} & 0.1089 & \textbf{0.2333}\, & 0.1608 & \textbf{0.3167} & 0.0934 & 0.1429 & 0.1119 & 0.1524\\
        \emph{HEmb\_Eq\ref{sim}} & 0.1111 & 0.2100 & 0.1640$*\dagger$ & 0.3133 & 0.0987 & \textbf{0.1524}\, & 0.1140$*$ & 0.1619\\
        \emph{WEmb\_Eq\ref{sim}} & \textbf{0.1137}$*$ & 0.2267 & \textbf{0.1654}$*\dagger$ & 0.3133 & \textbf{0.1012} & 0.1476 & \textbf{0.1154}$*$ & 0.1571\\
        \hline
    \end{tabular}
    \label{tab}
\end{table}
\vspace*{-5mm}

Table \ref{tab} shows that exploiting relations between concepts and using a relation-based similarity measure introduce noise, where the MAP of \emph{NoEmb-Leacock} is lower than the MAP of \emph{NoEmb-NoSim} for both IR models and in both collections. P@10 is also lower in \emph{clef11} and slightly better in \emph{clef12}.

\emph{WEmb} gives the best MAP among our approaches, where we use external statistical knowledge beside word embedding vectors. The comparison of \emph{WEmb} to \emph{NoEmb-NoSim} shows that representing concepts by vectors and using vector based inter-concept similarity improve the results. In 3 out of 4 cases the improvement is statistically significant. Moreover, there is no degradation in P@10. If we compare  \emph{WEmb} with \emph{NoEmb-Leacock}, we see that there is a small gain of MAP (for \emph{clef11} and \emph{bm25} the gain is statistically significant), and without corrupting P@10. Our approaches to represent concepts by vectors, and use vector-based similarity, improve MAP without corrupting P@10, i.e. the approaches are able to improve results without introducing noise. The only exception is \emph{HEmb}, where building concepts vectors considers the same word several times if it appears in several strings of the same concept, which represents a possible source of noise.


\section{Conclusion}

We presented three approaches to build concept embedding vectors based on pre-trained word embedding vectors. We used concepts vectors along with a vector-based similarity to improve IR performance. The results are promising, where the overall performance is improved without losing the absolute precision.

This study can be extended by achieving more in depth free parameters tuning, especially for vector size. Furthermore, we mainly compare the performance of a path-based measure, i.e. Leacock, to a vector-based measure (\ref{sim}) \cite{tpx2007}. However, we can also compare the results to content-based measures \cite{tpx2007}.

Both words and concepts are represented in the same vector space, so they are comparable. It is thus possible to compare concepts to the original textual content of documents. This is helpful to either achieve conceptual indexing or to improve the quality of some conceptual indexing methods like MetaMap by filtering out non-related or noisy concepts.

\vspace{-5mm}
\bibliographystyle{splncs03}
\bibliography{ref}

\end{document}